\documentclass[%
mathleft,%
]{an}
\usepackage{graphicx}
\usepackage{times}
\usepackage[abs]{overpic}

\def\Alfven{Alfv\'{e}n}
\def\Alfvenic{Alfv\'{e}nic}

\begin{document}

\Pagespan{1}{}
\Yearpublication{2012}%
\Yearsubmission{2012}%
\Month{1}%
\Volume{999}%
\Issue{92}%

\title{Evolution of solar-type stellar wind}

\author{Takeru K. Suzuki\inst{1}\fnmsep\thanks{Corresponding author:
  \email{stakeru@nagoya-u.jp}}
}
\titlerunning{Evolution of stellar wind}
\authorrunning{Takeru K. Suzuki}
\institute{
Department of Physics, Nagoya University, Furo-cho, Chikusa, Nagoya, Aichi, 
606-8602, Japan
}

\received{XXXX}
\accepted{XXXX}
\publonline{XXXX}


\abstract{%
By extending our self-consistent MHD simulations for the solar wind, 
we study the evolution of stellar winds of solar-type
stars from early main sequence stage to red giant phase. 
Young solar-type stars are active and the mass loss rates are larger by up 
to $\sim$ 100 times than that of the present-day sun. 
We investigate how the stellar wind is affected when 
the magnetic field strength and fluctuation amplitude at the photosphere 
increase. While the mass loss rate sensitively depends on the input energy from 
the surface because of the global instability related to the reflection
and nonlinear dissipation of \Alfven waves, it saturates 
at $\sim$ 100 times because most of the energy is used 
up for the radiative losses rather than the kinetic energy of the wind.
After the end of the main sequence phase when the stellar radius expands by 
$\sim$ 10 times, the steady hot corona with temperature $10^6$ K, suddenly 
disappears.  Chromospheric materials, with hot bubbles 
embedded owing to thermal instability, directly stream out; the red giant 
wind is not a steady stream but structured outflow.  
}

\maketitle

\section{Introduction}
The origin of the energy which drives the solar wind is believed to 
be in the surface convection zone. Turbulent motion associated with 
the convection excites various modes of waves that propagate upwardly. 
Magnetic field lines, which are also a consequence of dynamo processes 
in the surface convective motion (e.g. Yoshimura 1975; Parker 1993), play 
as a guide for the upgoing waves. 
Among various types of waves, the \Alfven wave is the most reliable candidate 
which transfers the energy from the surface convection to the upper region 
where the solar wind is accelerated because it can travel a longer 
distance due to the incompressive nature (e.g. Alazraki \& Couturier 1971; 
Belcher 1971; Hollweg 1973). 

\Alfven waves have been extensively studied in the context of driving 
the solar wind. The dissipation of \Alfven \\waves is a key to understand 
the acceleration of the solar wind because it controls the energy and momentum 
transfer from the waves to the gas. Various types of dissipation mechanisms 
have been investigated such as ion-cyclotron resonance (e.g. Axford 
\& McKenzie 1992; Kohl et al.1998), turbulent cascade (e.g. Matthaeus et 
al.1996; Verdini \& Velli 2007), 
and the nonlinear mode conversion to compressive modes (e.g. Kudoh \& Shibata 
1999) by using steady-state models (e.g. Cranmer \& van 
Ballegooijen 2007) or dynamical simulations (e.g. Suzuki \& Inutsuka 2005). 

Among others, we have carried out magnetohydrodynamical (MHD) simulations 
covering from the photosphere to the sufficiently outer region where the 
solar wind is already accelerated  (Suzuki \& Inutsuka 2005, 2006; 
Matsumoto \& Suzuki 2012). One of the greatest advantage of our dynamical 
simulations is that the inner boundary is not an ad hoc `coronal base'
or `upper chromosphere' but the photosphere where there are much observational 
information\footnote{Recent steady-state calculations (e.g. Verdini \& Velli 
2007; Cranmer \& van Ballegooijen 2007) also include the photosphere.} 
(e.g. Fujimura \& Tsuneta 2009; Matsumoto \& Kitai 2010). 
Therefore, mass loss rate can be directly determined as 
an output of the surface properties.  

This type of wind is not unique to the solar wind. 
Main sequence stars with mass comparable to or less than the solar mass 
possess surface convection zones, and the stellar winds emanating from 
these stars by similar mechanism (e.g. Cranmer \& Saar 2011).
Observations of asterospheres of low-mass to intermediate-mass stars show 
that the mass loss rates of young stars are much larger up to $\sim$ 100
times of the present solar value (Wood et al. 2002, 2005).   

Red giant stars also have surface convection zones. \Alfven \\waves possibly 
play a role in driving the stellar winds from red giant stars (e.g. Hartmann 
\& MacGregor 1980; Holzer et al.1983).

In this paper, we discuss stellar winds originating from surface 
convection zones. In particular we focus on how the mass loss rate is 
determined from the photospheric properties, based on our MHD simulations.

\section{Our previous works for the present-day solar wind}
In this section, we briefly summarize our MHD simulations for the solar wind.
In Suzuki \& Inutsuka (2005, 2006), we performed self-consistent 
one-dimensional (1D) simulations that handle the propagation, reflection, 
and dissipation of MHD waves in an individual super-radially open magnetic 
flux tube from the photosphere to the solar wind region. 
We took into account radiative cooling and thermal conduction to 
examine the coronal heating without ad hoc assumptions and three 
components of magnetic field and velocity to treat \Alfven waves. 
We injected transverse fluctuations from the photosphere 
and run the simulations until the quasi-steady states were achieved. 
A main result of these papers is that the input of the transverse fluctuations 
with $\sim$ 1 km/s at the photosphere naturally drives the solar wind which 
is observed today. 

The \Alfven waves which are excited from the photosphere are mostly 
reflected back downward because of the change of the \Alfven speed (Moore 
1991), but roughly 10 \% of the initial Poynting flux associated 
with the \Alfven waves penetrates to the corona and contributes to 
the heating of the solar wind. The main channel of the 
dissipation of the \Alfven \\waves is the generation of compressive 
waves, particularly slow MHD waves (Kudoh \& Shibata 1999). 
The steepening of the wave fronts leads to the shock dissipation of 
these excited compressive waves. 
In general, however, in the 1D simulations the shock dissipation tends to be 
overestimated because the waves are confined in the individual flux tubes 
without leakage. 

In Matstumoto \& Suzuki (2012), we extended to two dimensional (2D) 
simulations. The biggest difference is that we can treat the leakage 
of the waves to neighboring flux tubes. Cascading \Alfvenic turbulence 
can be also handled, although it is restricted to the 2D space. 
In the 2D simulation, the effect of compressive waves is suppressed, 
while the turbulent cascade becomes comparably important in the dissipation 
of the \Alfven waves. The overall dissipation rate of the \Alfven waves 
is comparable to that of the 1D cases, and as a result, the 1D and 2D 
simulations give  
similar mass loss rates and terminal velocities.

In the following sections, we extend these simulations from the present-day  
solar wind to more active solar-type stars (\S 3) and red giant stars (\S 4). 
Because the 2D and 1D simulations yield similar global properties of 
the winds, we use the 1D simulations to save computational time.

\section{Mass loss from young active solar-type stars}
\begin{figure}[h]
\includegraphics[width=0.8\linewidth,height=50mm]{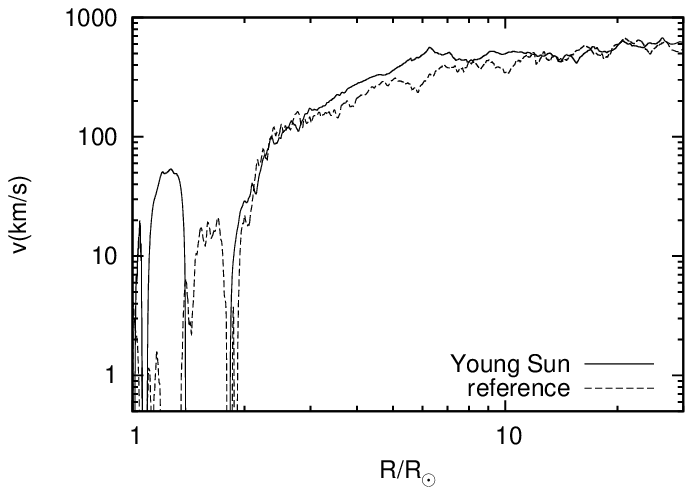}
\includegraphics[width=0.8\linewidth,height=50mm]{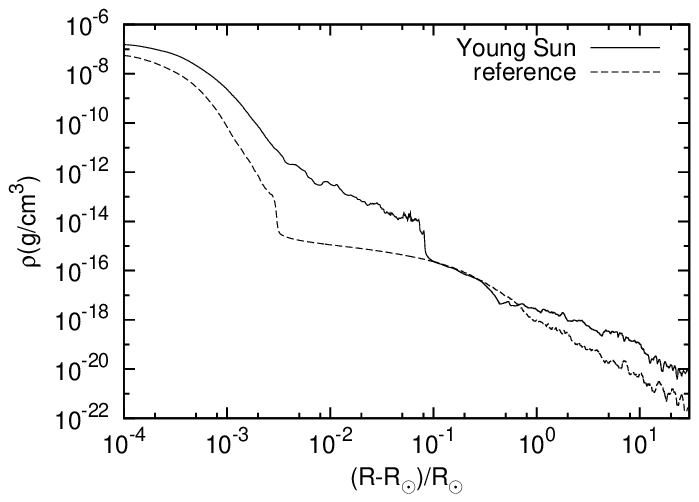}
\includegraphics[width=0.8\linewidth,height=50mm]{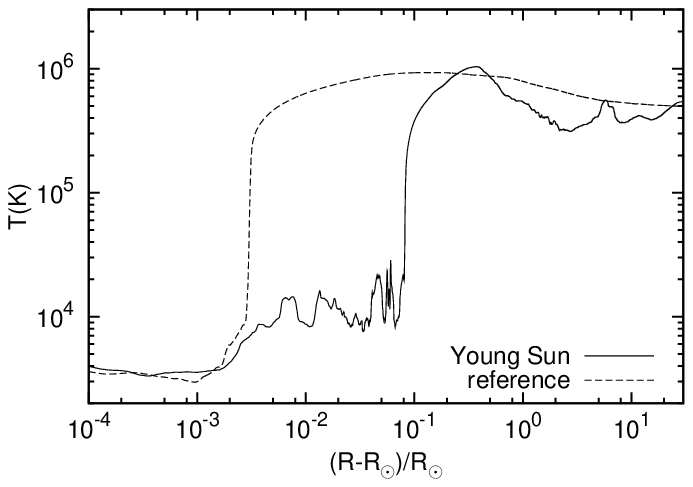}
\caption{Comparison of the wind structures. The solid lines 
are the wind structure for the active case (see text) and the dashed lines 
are the wind structure of the reference case. From top to bottom, velocity, 
density, and temperature are displayed.}
\label{fig:wndstr}
\end{figure}

Young solar-type stars are generally very active. 
The X-ray flux is up to $\sim$ 1000 times larger than the present Sun 
(e.g. G\"{u}del 2004), and the transition region flux is also much 
larger (Ayres 1997). 
Observation of young main sequence stars show very strong magnetic 
fields with an order of kG or even larger 
(e.g., Donati \& Collier Cameron 1997; Saar \& Brandenburg 1999). 
The mass loss rate is also much higher but seems saturated around 
$\sim$ 100 times of the present solar level (Wood et al. 2002; 2005).

\begin{figure*}[t]
  \begin{overpic}[height=50mm,width=0.4\linewidth]{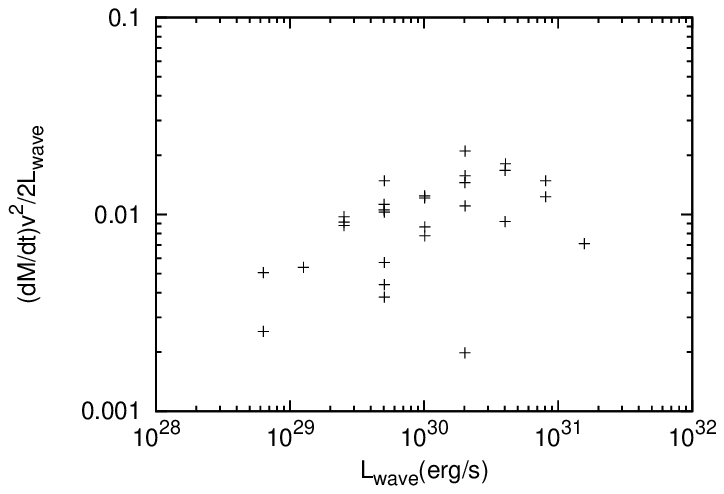}
    \put(60,70){\vector(3,2){55}}
    \put(140,110){\vector(3,-2){20}}
  \end{overpic}
  \begin{overpic}[height=50mm,width=0.4\linewidth]{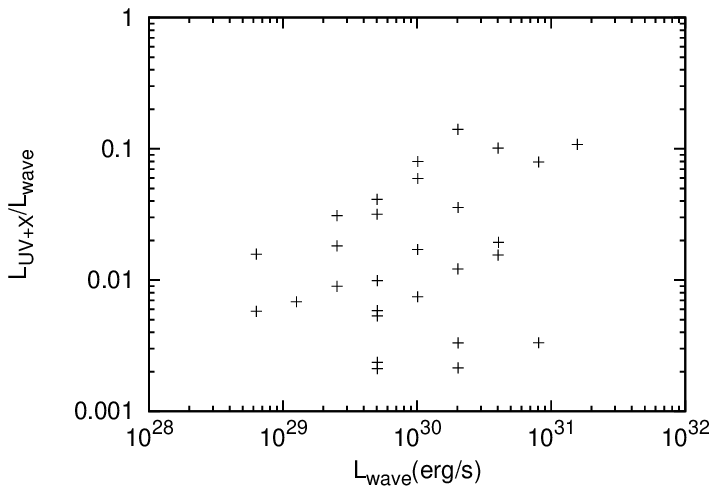}
    \put(60,70){\vector(3,2){60}}
  \end{overpic}
\caption{Energetics of stellar wind derived from the simulations; each 
data point indicates each run.  
The left panel shows the fractions of the kinetic energy fluxes on the 
total input Poynting fluxes from the photospheres, and the right panel shows 
the fractions of the total radiative losses from the transition regions and 
coronae on the total input Poynting fluxes. }
\label{fig:Lwdep}
\end{figure*}

In this section, we investigate how the solar atmosphere reacts to increases 
of magnetic field strength and fluctuation amplitude at the photoshere. 
As the reference case we set up a magnetic flux tube with field strength 
$B_0=1$ kG at the photosphere and a super-radial expansion factor $f = 1000$ 
to match recent HINODE observations (Tsuneta et al. 2009; Ito et al. 2010), 
and inject fluctuations, $\delta v=1.4$ km s$^{-1}$ from the photosphere. 
This reference case well reproduces 
the average global properties of the present-day solar wind. 
We perform more than 30 simulation runs with different $B_0$, $f$, and 
$\delta v$.     

Figure \ref{fig:wndstr} shows the stellar wind structure which adopts 
$B_0=2$ kG, $\delta v = 2.8$ km s$^{-1}$, and $f=1000$ (solid lines, labeled as 
`Young Sun') in comparison with the reference case (dashed lines). 
This active case gives the mass loss rate, $\dot{M} = 4\times 10^{-13}$ 
$M_{\odot}$/yr, which is 20 times larger than $2\times 10^{-14}$ $M_{\odot}$/yr 
obtained from the reference case ($\approx$ the present value). 
Since the injected Poynting flux ($\propto B_0 \delta v^2$) in the active 
case is only 8 times larger, the mass loss rate quite sensitively depends on 
the energy input. This is because of the global instability involved with 
the reflection and nonlinear dissipation of \Alfven waves 
(see Suzuki 2012 for details). 

The location of the transition region in the active case is at 
$\approx 1.1 R_{\odot}$, which is much higher than the height of the
transition region of the reference case ($\approx 1.003 R_{\odot}$). 
This is because in the active case more material is lifted up due to 
the larger energy injection from the photosphere and the temperature 
cannot increase easily owing to the larger radiative cooling. Hence, 
the chromosphere extends to higher altitude in the active case. 
Interestingly enough, the observation of a young solar-like star, CoRot-2a, 
by using the Rossiter-McLaughlin effect shows that this star possesses an
extended chromosphere of up to 1.16 times of the stellar radius (Czesla et al.
2012). 
 
Figure \ref{fig:Lwdep} displays the fractions of the kinetic energy 
flux of the stellar wind, $\dot{M}\frac{v^2}{2}$, (left) and the radiative 
loss from the transition 
region and corona (right) of each run against the total injected Poynting 
energy associated with \Alfven\\ waves, $L_{\rm wave}$, from the photosphere. 
Both panels show large scatters in the vertical direction, reflecting 
varieties of $f$. 
The sums of these two fractions are much less than unity in all the 
runs, because most of the input energies are reflected back downward. 

The left panel shows that the fraction of the kinetic energy flux initially 
increases with increasing input energy; the increase of the kinetic energy of 
the stellar wind is more rapid than linear on the input energy, which is 
related with the global instability owing to the reflection and nonlinear 
dissipation of \Alfven waves as explained 
above (Suzuki 2012). However, it eventually saturates and decreases on 
increasing $L_{\rm wave}$. On the other hand, the radiative loss 
shows saturation as well but the overturning trend is not so distinctive.

\begin{figure}[h]
  \begin{overpic}[height=55mm,width=0.9\linewidth]{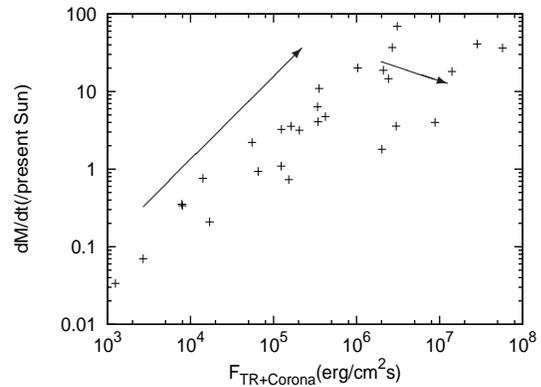}
      \put(60,75){\vector(1,1){60}}
      \put(150,130){\vector(3,-1){25}}
    \end{overpic}
\caption{Mass loss rate normalized by the present solar value on 
radiative flux from transition region and corona.}
\label{fig:LMdot}
\end{figure}

Combining these two panels, we can get the relation between radiative flux 
and mass loss rate 
(Fig. \ref{fig:LMdot}), which can be directly compared with the figures 
presented in Wood et al. (2002, 2005). 
As expected from Fig. \ref{fig:Lwdep}, the mass loss rate is positively 
correlated with radiative flux initially, while it
eventually saturates, or even slightly decreases.  The main reason 
of the saturation is that most of the input energies are exhausted for 
the radiative losses besides reflection in these saturated cases. 
Namely, the increase of $\dot{M}$ is mainly done by the increase of 
density, which enhances radiative losses. Finally, no more energy remains
for the kinetic energy of the stellar wind. We discuss this saturation 
mechanism in more detail in a forthcoming paper (Suzuki et al. 2012, in 
preparation). 

\section{Mass loss from red giants}

\begin{figure}[h]
\includegraphics[width=0.8\linewidth,height=96mm]{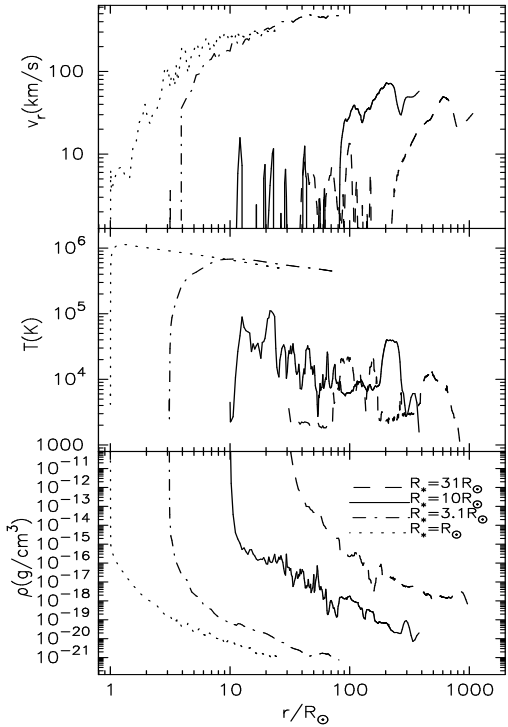}
\caption{From the top to the 
bottom, radial outflow velocity, $v_r$ (km s$^{-1}$), temperature, $T$ (K),  
and density, $\rho$(g cm$^{-3}$), are plotted. 
The dotted, dash-dotted, solid, and dashed lines are the 
results of stellar radii, $R=R_{\odot}$ (the present Sun), $3.1R_{\odot}$ 
(sub-giant), $10R_{\odot}$ (red giant), and $31R_{\odot}$ (red giant), 
respectively.}
\label{fig:rgbwnd}
\end{figure}

After the hydrogen in the core is exhausted, a star evolves 
to a red giant stage. Because red giant stars have surface convection 
zones, it is expected that the fluctuations the surfaces would give a 
significant contribution to the stellar winds. 
On the other hand, the expansion of the 
radius largely affects the dynamics through the change of the surface gravity. 
We have carried out MHD simulations for red giant winds (Suzuki 2007), 
of which we briefly summarize the results. 

Figure \ref{fig:rgbwnd} presents the evolution of stellar winds of 
a $1M_{\odot}$ star from main sequence to red giant stages. 
The surface amplitude is estimated from the convective flux (Renzini et al.
1977), whereas this should be tested by comparison with recently observed
chromospheric fluxes (e.g. P\'{e}rez Martin\'{e}z et al. 2011). 
The middle panel shows that the average temperature drops suddenly 
from $T\simeq 7\times 10^5$K in the sub-giant star (dash-dotted) 
to $T\le 10^5$K in the red giant stars, which is consistent with the observed 
``dividing line'' (Linsky \& Haisch 1979). 
The main reason of the disappearance of the steady hot coronae is that 
the sound speed ($\approx 150$ km s$^{-1}$) 
of $\approx 10^6$ K plasma exceeds the escape speed, 
$v _{\rm esc}(r)=\sqrt{2G M_{\star}/r}$, at a few stellar radii 
in the red giant stars; 
the hot corona cannot be confined by the gravity any more in the atmospheres 
of the red giant stars.  
Therefore, the material flows out before heated up to coronal temperature 
as an extended chromosphere (Schr\"{o}der \& Cuntz 2005) with intermittent 
hot bubbles (Suzuki 2007; Crowley et al.2008).  
In addition, the thermal instability of the radiative cooling function 
(Landini \& Monsignori-Fossi 1990)
plays a role in the sudden decrease of temperature; since the gas with $T=
10^5 - 10^6$ K is unstable, the temperature quickly decreases from 
the subgiant 
to red giants. 

The densities of the winds increase with stellar evolution due to the decrease 
of the surface gravity. 
The wind velocities are much smaller than the 
surface escape speeds, because the onset locations of the winds are around 
several stellar radii (Harper et al.2009). Through the stellar evolution,  
the mass loss rate increases due to 
the increase of the stellar surface ($\propto R^2$) and the increase of 
the density. From the main sequence star to the read giant star with 
$R_{\star}=31R_{\odot}$, the mass loss rate increases $10^5 - 10^6$ times.

\acknowledgements
This work was supported in part by Grants-in-Aid for 
 Scientific Research from the MEXT of Japan, 22864006.

\end{document}